\documentclass[aps,prb,twocolumn,flushbottom,twoside,superscriptaddress]{revtex4}
\usepackage{graphicx}
\usepackage{amsmath}

\begin{document}

\title{Theory of Half-Metallic Double Perovskites II: Effective Spin Hamiltonian and Disorder Effects}
\author{Onur Erten}
\affiliation{Department of Physics, The Ohio State University, Columbus, Ohio 43210, USA}
\author{O. Nganba Meetei}
\affiliation{Department of Physics, The Ohio State University, Columbus, Ohio 43210, USA}
\author{Anamitra Mukherjee}
\affiliation{Department of Physics, The Ohio State University, Columbus, Ohio 43210, USA}
\affiliation{Department of Physics and Astronomy, University of British Columbia, Vancouver, BC V6T 1Z1, Canada}
\author{Mohit Randeria}
\affiliation{Department of Physics, The Ohio State University, Columbus, Ohio 43210, USA}
\author{Nandini Trivedi}
\affiliation{Department of Physics, The Ohio State University, Columbus, Ohio 43210, USA}
\author{Patrick Woodward}
\affiliation{Department of Chemistry, The Ohio State University, Columbus, Ohio 43210, USA}

\begin{abstract}
Double perovskites like Sr$_2$FeMoO$_6$ are materials 
with half-metallic ground states and ferrimagnetic T$_{\rm{c}}$'s well above room temperature. 
This paper is the second of our comprehensive theory for half metallic double perovskites.
Here we derive an effective Hamiltonian for the Fe core spins by ``integrating out'' the itinerant Mo electrons
and obtain an unusual double square-root form of the spin-spin interaction.
We validate the classical spin Hamiltonian by comparing its results with those of the full quantum treatment presented in 
the companion paper ``Theory of Half-Metallic Double Perovskites I: Double Exchange Mechanism''. We
then use the effective Hamiltonian to compute magnetic properties as a function of temperature and disorder
and discuss the effect of excess Mo, excess Fe, and anti-site disorder on the magnetization and T$_{\rm{c}}$.
We conclude with a proposal to increase T$_{\rm{c}}$ without sacrificing carrier polarization.

\end{abstract}

\maketitle
Strong electron correlations and the interplay among charge, spin and lattice degrees of freedom lead to a wide range of 
spectacular phenomena in transition metal oxides\cite{Imada} such as high T$_{\rm{c}}$ superconductivity, colossal 
magnetoresistance and large thermopower. 

Half metals with fully spin polarized ground states provide another example of such unique and spectacular phenomena.
Among the known examples, double perovskites (DPs) are of particular interest due to 
their high ferromagnetic $T_{\rm{c}}$'s along with the possibility of integrating different functionalities with oxide electronics\cite{Ibarra_2007}. 
One of the best-studied half-metallic DP is Sr$_2$FeMoO$_6$ 
(SFMO) with $T_{\rm{c}}$=420 K, well above room temperature\cite{Ibarra_2007,Kobayashi_1998,Sarma_2000}. 
DPs have the form A$_2$BB$^\prime$O$_6$ , which
is derived from the simple ABO$_3$ perovskite structure with a three-dimensional (3D) checkerboard ordering of B and B$^\prime$ 
ions. DPs have a range of fascinating properties from spin liquids to multiferroics, as well as from metals to 
multi-band Mott insulators.\cite{Ibarra_2007, Kobayashi_1998, Balents_DP1, Balents_DP2, nganba_2012}

This article is the second part of our comprehensive theory for half metallic double perovskites.
Along with its companion paper titled ``Theory of Half-Metallic Double Perovskites I: Double Exchange Mechanism''\cite{our_PRB1} (hereafter referred to as paper I), it is an extension of our recent Letter \cite{our_PRL}. We begin by summarizing the first paper where 
we discuss the full quantum Hamiltonian describing core spins on Fe coupled to conduction 
electrons through a generalized double exchange mechanism. We calculated the magnetic and electronic properties as a function 
of temperature using exact diagonalization of the ``fast" electronic degrees coupled to ``slow" core spin configurations generated by classical 
Monte Carlo simulations (ED+MC). By retaining the electronic degrees of freedom, we obtained information about the temperature dependent density of states
and the destruction of the fully polarized half-metallic ground state through thermal fluctuations. One of our central results is that the
conduction electron polarization at the chemical potential is directly proportional to the core-spin magnetization.
This finding is significant because it indicates that if one can derive an effective Hamiltonian for the core spins, 
it would be possible to deduce the electronic polarization, a quantity of central importance for spin injection and spin transport, but one that is difficult to measure directly. 
The effective Hamiltonian also has the advantage that it can be used to simulate large system sizes compared to severe size limitations faced by ED+MC methods.
\begin{figure}[!t]
\includegraphics[width=8cm]{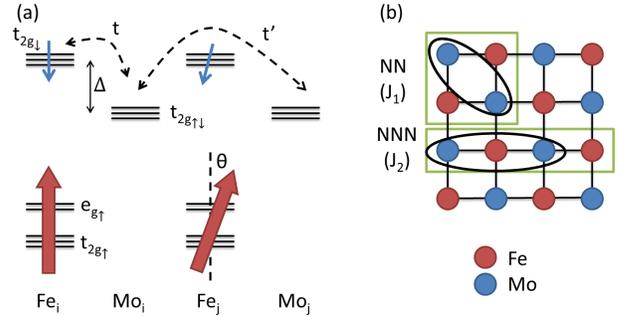}
\caption{(a) Schematic showing energy levels at transition metal sites in
two unit cells (formula units) of SFMO. The Fe sites have localized $S=5/2$ core spins, treated as classical
vectors with orientation $(\theta,\phi)$. The parameters $t,t^\prime$ and $\Delta$
of the Hamiltonian, Eq. \eqref{eq:quantum_hamiltonian}, 
governing the dynamics of the itinerant electrons in t$_{2g}$ orbitals, are also shown. (b) Nearest neighbor (NN) and 
next nearest neighbor (NNN) configuration of two unit cells of DPs. 
\label{Fig:1}}
\end{figure}

With this motivation, here we focus on describing the properties of the Fe core spins by ``integrating out'' the itinerant Mo electrons. The main 
results are: (1) We derive a new effective Hamiltonian, H$_{\rm eff}$, for the classical spins by generalizing, in a non-trivial way, 
the Anderson-Hasegawa analysis for manganites\cite{Anderson_1955} to double perovskites. The functional form of H$_{\rm eff}$ is different from standard
Heisenberg or Anderson-Hasegawa Hamiltonians. (2) We validate H$_{\rm{eff}}$ by comparing its spin wave
dispersion and temperature-dependent magnetization M(T) with that of the full Hamiltonian obtained from the ED+MC method.
H$_{\rm eff}$ indeed captures the magnetic properties of the full Hamiltonian at all temperatures whereas the Heisenberg Hamiltonian can only describe
the low temperature behavior. (3) We have performed the first 3D finite temperature calculations of magnetic properties of DPs with accurate estimates
of T$_{\rm c}$ using finite size scaling. (4) The effective Hamiltonian also allows us to efficiently study the effects of disorder on M(T). While 
both excess Fe and Mo decrease the saturation magnetization and T$_{\rm{c}}$, anti-site disorder in which Fe and Mo exchange places, behaves 
differently; although magnetization drops, T$_{\rm{c}}$ is not affected. (5) The previous result forms the basis of our proposal to increase T$_{\rm{c}}$ 
without sacrificing conduction electron polarization. We propose that by putting 
excess Fe and compensating the loss of carriers with La doping can indeed lead to a dramatic increase in T$_{\rm{c}}$. 

We start by briefly describing the full quantum Hamiltonian.
We then solve the problem of two unit cells and derive the effective exchange Hamiltonian 
between two Fe core spins and generalize this form to the infinite lattice.

For large Hund's coupling J$_{\rm{H}}$, Fe$^{3+}$ in the 3d$^5$ configuration saturates the ``up'' manifold and forms a large
spin {\bf{S}}=$5/2$ that we treat classically with a local axis of quantization 
along ${\bf S}_i$. Mo$^{5+}$ (4d$^1$) contributes to conduction in t$_{\rm{2g}}$ orbitals. 
Due to the symmetry of t$_{\rm{2g}}$ orbitals, d$_{\alpha \beta}$ orbitals can only delocalize in $\alpha \beta$ 
planes\cite{harris_2004} ($\alpha \beta=xy,yz,xz$). For all the Mo sites $j$, we choose the same (global) axis of quantization.
The generalized double exchange Hamiltonian\cite{alonso_2003,sanyal_2009, our_PRL, Aligia_2001, Petrone_2002}
that describes the core spins interacting with conduction electrons is
\begin{eqnarray}
H&=& -t\sum_{\langle i,j\rangle,\sigma}
(\epsilon_{i\sigma}d^{\dagger}_{i\downarrow}c_{j\sigma}+h.c.)
\cr
&&
-t^{\prime}\sum_{\langle j,j^\prime \rangle,\sigma}c^{\dagger}_{j\sigma}c_{j^\prime \sigma}
+ \Delta \sum_{i}d^{\dagger}_{i\downarrow}d_{i\downarrow}
\label{eq:quantum_hamiltonian}
\end{eqnarray}
where $d_{i\sigma} \, (c_{i\sigma})$ are fermion
operators on the Fe (Mo) sites with spin $\sigma$.

The orientation $(\theta_i,\phi_i)$ of the classical spins ${\bf S}_i$
affects the Mo-Fe hopping via
$\epsilon_{i \uparrow}= - \sin(\theta_i/2)\exp(i\phi_i/2)$ and
$\epsilon_{i \downarrow}= \cos(\theta_i/2)\exp(-i\phi_i/2)$. 
\section{Exact Solution of Two Site Problem}
We solve the Hamiltonian in Eq. \eqref{eq:quantum_hamiltonian} exactly analytically for two unit cells, shown schematically in Fig.~\ref{Fig:1}. 
This is a generalization of the Anderson and Hasegawa analysis for manganites\cite{Anderson_1955} applied to double perovskites. 

In a single unit cell, there are three states derived from the Fe$_{\downarrow}$
and Mo$_{\uparrow, \downarrow}$ t$_{\rm{2g}}$ orbitals. We label the unit cells as i and j, and without loss of generality choose a
coordinate system such that one of the core spins $\textbf{S}_i$ 
is aligned with the z axis, and the other core spin $\textbf{S}_j$ lies in the x-z plane (Fig.~\ref{Fig:1}(a)). 
This particular choice of coordinates simplifies the calculation as it gauges away the
$\phi$ dependence. Thus,  $\epsilon_{\uparrow} =\sin(\theta_i/2)$  and $\epsilon_{\downarrow} =\cos(\theta_i/2)$ , 
where $\theta$ is the relative angle between $\textbf{S}_i$ and $\textbf{S}_j$. 
The two unit cell Hamiltonian is given by

\begin{widetext}
\begin{eqnarray}
 H =
 \left( {\begin{array}{cccccc}
  \Delta & 0 & -t & 0 & 0 & -\gamma t \\
  0 & 0 & 0 & -\sin(\theta/2)t & 0 & 0\\
  -t & 0 & 0 & -\cos(\theta/2)t & 0 & 0 \\
  0 & -\sin(\theta/2)t & -\cos(\theta/2)t & \Delta & t\sin(\theta/2) & -t\cos(\theta/2) \\
  0 & 0& 0& t\sin(\theta/2) & 0 & 0 \\
  -\gamma t & 0 & 0 &  -\cos(\theta/2)t &0 & 0 \\
 \end{array} } \right)
\label{eq_matrix}
\end{eqnarray}
\end{widetext}
in the basis of \{Fe$_{{\rm i}\downarrow}$, Mo$_{{\rm i}\uparrow}$, Mo$_{{\rm i}\downarrow}$, Fe$_{{\rm j}\downarrow}$, 
Mo$_{{\rm j}\uparrow}$, Mo$_{{\rm j}\downarrow}$\}.
Here $\gamma$=1 for nearest neighbor (NN) and 0 for  next nearest neighbor (NNN) configurations (See Fig. \ref{Fig:1}(b)). 
By converting the 6$\times$6 matrix for H in a block diagonal form, it can be solved analytically. 
The eigenvalues are only a function of the angle between the core spins {\bf S$_{\rm i}$} and {\bf S$_{\rm j}$} and describe the 
effective magnetic exchange Hamiltonians.
\begin{figure}[!t]
\includegraphics[width=5cm]{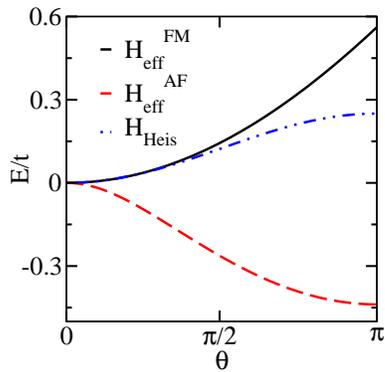}
\caption{Energy as a function of $\theta$ for ferromagnetic H$_{\rm{eff}}$ (one electron in two unit cells) and antiferromagnetic 
H$_{\rm{eff}}$ (two electrons in two unit cells) . 
Effective Hamiltonian gives hints for filling dependent magnetic phase transition. 
We include FM Heisenberg Hamiltonian (H$_{\rm{Heis}}$)
for comparison. Note that FM H$_{\rm{eff}}$ is quadratic for a broader range of $\theta$ compared to H$_{\rm{Heis}}$.
\label{Fig:2}}
\end{figure}
For the nearest neighbor configuration ($\gamma=1$), the lowest eigenvalue, describing one electron in two unit cells which 
corresponds to an electronic density of $n=0.5$,  is
\begin{eqnarray}
 H^{\rm{FM}}_{\rm{eff}}=-\sqrt{(\Delta/2)^2+2t^2(1+\cos(\theta/2))} 
\end{eqnarray}
or equivalently
\begin{eqnarray}
H^{\rm{FM}}_{\rm{eff}}=-\sqrt{(\Delta/2)^2+2t^2(1+\sqrt{(1+\textbf{S}
_i\cdot\textbf{S}_j)/2} ~)}
\label{eq5}
\end{eqnarray}
where $\textbf{S}$ is the unit spin vector. 
We obtain a very interesting modified functional form with a
double square root structure that is different from conventional Heisenberg or previously studied
Anderson-Hasegawa models\cite{Anderson_1955}. Note that the interaction is ferromagnetic with spin stiffness $J_{FM} \equiv\partial^2 E/ \partial \theta^2$ 
obtained by expanding the energy close to  $\theta=0$,
where $E(\theta)\approx E (0)+(1/2)(\partial^2 E/ \partial \theta^2) \theta^2+{\cal{O}} (\theta ^4)$. We find, 
\begin{eqnarray}
J_{FM}\sim  \left\{
\begin{array}{ll}
-t & {\rm{for}} {~} t\gg |\Delta| \\
-t^2/\Delta, & {\rm{for}} {~} t\ll |\Delta| 
\end{array}
\right.
\label{eq6}
\end{eqnarray}
showing that the kinetic energy of the conduction electrons sets the scale of the ferromagnetic exchange. 

\begin{figure}[!t]
\includegraphics[width=8cm]{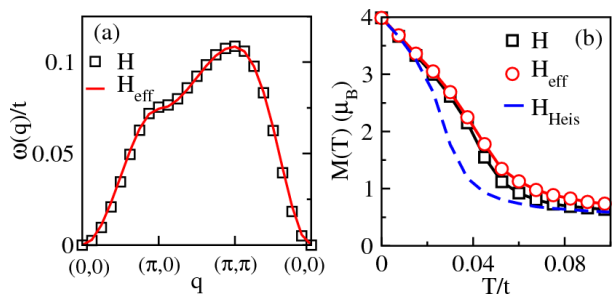}
\caption{(a) Spin wave spectrum of full Hamiltonian and the H$_{\rm{eff}}$, (b) M(T) comparison between full Hamiltonian, 
H$_{\rm{eff}}$, and Heisenberg Hamiltonian. All simulations are done with an 8$\times$8 system due to the high
computational cost of the exact diagonalization and Monte Carlo calculations.
\label{Fig:3}}
\end{figure}

For two electrons in two unit cells, which corresponds to $n=1$, the effective Hamiltonian is obtained by 
adding up the lowest two eigenvalues. For the NN configuration, the effective Hamiltonian is antiferromagnetic given by
\begin{eqnarray} 
H^{\rm{AF}}_{\rm{eff}}&=& -\sqrt{(\Delta/2)^2+2t^2(1+\cos(\theta/2))}\cr
&&
-\sqrt{(\Delta/2)^2+2t^2(1-\cos(\theta/2))}.
\label{eq7}
\end{eqnarray}

Upon increasing the electron density ($n=0.5 \rightarrow 1$), we find that the effective magnetic coupling changes from ferromagnetic
to antiferromagnetic which is rather unconventional. Metallic antiferromagnetism with large local 
moments at a commensurate wave vector is rare in nature. Even at a two unit cell level, H$_{\rm{eff}}$ provides a hint for this transition and illuminates the mechanism, though only discrete fillings 
are accessible at this level. The filling driven FM-AFM transition has also been discussed by others\cite{sanyal_2009}.
 The exchange stiffness is given by,
\begin{eqnarray}
J_{AF}\sim \left\{
\begin{array}{ll}
t & {\rm{for}} {~} t\gg |\Delta| \\
t^4/|\Delta^3|, & {\rm{for}} {~} t\ll |\Delta| 
\end{array}
\right.
\end{eqnarray}
with the scale for antiferromagnetism also set by the kinetic energy of the conduction electrons. 

As we will discuss in the following section, SFMO with a conduction electron density n=0.33 is far from any antiferromagnetic instability.
We therefore consider only the ferromagnetic form of the two spin interaction.
For convenience, we define two functions $F_1(x)$ and $F_2(x)$ that capture the NN and NNN ferromagnetic interactions respectively:
\begin{equation}\label{f1}
 F_{1}(x)=8\sqrt{2+\sqrt{2 + 2x}}
\end{equation}
and 
\begin{equation}\label{f2}
 F_{2}(x)=(5+\sqrt{5})\sqrt{6+2\sqrt{3+2x}}
\end{equation}

where $x=\mathbf{S}_i \cdot \mathbf{S}_j$.
Up to a constant factor of 8, $F_{1}(x)$ is obtained by setting $\Delta=0$ (see Appendix) in Eq. \ref{eq5}.
A similar procedure for the NNN exchange with $\gamma$ set to zero in Eq.~\ref{eq_matrix} yields $F_2(x)$.

\section{Effective Spin Hamiltonian}
Here we extend the analysis of {\em ferromagnetic} two spin interaction discussed in the previous section to a full lattice in order to study the magnetic properties of SFMO. The effective spin Hamiltonian with NN and NNN interactions has the following form 
\begin{equation}\label{eq:H_eff}
 H_{\rm eff}= {-J_1}\sum_{\langle i,j
\rangle}F_{1}\left(\mathbf{S}_i\cdot\mathbf{S}_j\right) -
J_2\sum_{\langle\langle i,j
\rangle\rangle}F_{2}\left(\mathbf{S}_i\cdot\mathbf{S}_j\right)
\end{equation}
where $F_{1(2)}(x)$ is defined in Eq.~\ref{f1}(\ref{f2}) and $x=\mathbf{S}_i\cdot \mathbf{S}_j$. 

We justify this Hamiltonian in two steps. First, we fix the values of $J_1$ and $J_2$ by matching the spin wave dispersion of H$_{\rm eff}$ with that of the full quantum Hamiltonian H (Eq.~\ref{eq:quantum_hamiltonian}). In the second step, we compare the magnetization as a function of temperature, $M(T)$, obtained from H$_{\rm eff}$ and H. In the details described below, we show that our effective Hamiltonian completely describes the magnetic properties of SFMO at all temperatures.  

For small $\theta$, $F_{1(2)}(\cos\theta)\approx {\rm const.} + (1/2)\theta^2$ which is the same as that of the Heisenberg interaction. The particular choice of prefactors (8 for $F_1$ and $(5+\sqrt{5})$ for $F_2$) allows this simple comparison. It is therefore not suprising that the spin wave spectrum  obtained by expanding H$_{\rm eff}$ around the FM ground state for small angle deviations is the same as Heisenberg model with NN and NNN interactions. As shown in Fig. \ref{Fig:3}(a), we can match the spin wave spectrum of full quantum Hamiltonian with that of H$_{\rm eff}$ by tuning J$_1$ and J$_2$. This gives us the required values of $J_1$ and $J_2$ in our model. The agreement over the entire spectral range, rather than just at small energies, is indeed remarkable. We also point out that for the full quantum H we have used $\Delta$=2.5$t$ and $t^\prime$=0.1$t$, however the effective spin-Hamiltonian is relatively 
insensitive to the value of $\Delta$ (see Appendix) and at the level of spin waves, the effects of $\Delta$ and $t^\prime$ are captured through $J_1$ and $J_2$. 
This justifies the simplifying assumption of $\Delta=0$ used to obtain $F_{1(2)}(x)$.  

For the second step of validating H$_{\rm eff}$, we perform ED+MC calculations for the full quantum Hamiltonian along with classical Monte Carlo simulations for H$_{\rm{eff}}$ and for the Heisenberg Hamiltonian. In Fig. \ref{Fig:3}(b) we present a comparison of the temperature-dependent magnetization M(T) calculated for each of these three Hamiltonians on an 8$\times$8 system. It is remarkable to observe that M(T) calculated from H$_{\rm{eff}}$ agrees remarkably well with that obtained for the full Hamiltonian 
{\em at all temperatures} thereby validating H$_{\rm{eff}}$. Note that the Heisenberg Hamiltonian is only able to explain M(T) at low temperatures
and fails at intermediate temperatures T$\simeq$T$_{\rm{c}}/2$.

Note that the full quantum model with classical spins coupled to conduction electrons has low-lying fermionic excitations. From 
a functional integral description, integrating out the fermions would give rise to various extra exchange 
terms like longer range interactions and four or more spin exchanges. 
The fact that we can reproduce M(T) using H$_{\rm{eff}}$ at all temperatures shows that the effect of such terms is negligible and we indeed capture the 
most important magnetic exchange interactions within our model. 

The agreement between M(T) for H$_{\rm{eff}}$ and the full quantum Hamiltonian also indicates that
both J$_1$ and J$_2$ are temperature independent. Although it is not clear {\it{a priori}} why this is the case, the fact that 
T$_{\rm{c}}$ is much less than the bandwidth provides a reasonable justification for the temperature independence of the exchange constants 
up to temperatures of 
order T$_{\rm{c}}$.
\begin{figure}[!t]
\includegraphics[width=8cm]{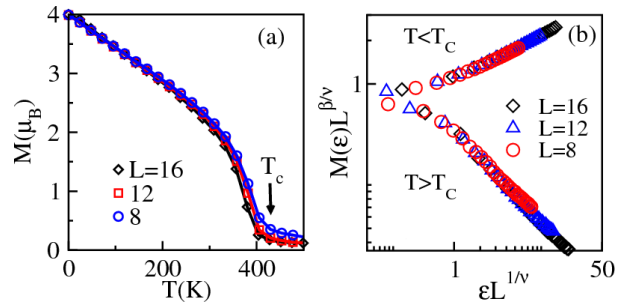}
\caption{ (a) Magnetization as a function of temperature, M(T), of H$_{\rm{eff}}$ by classical Monte Carlo calculations 
for increasing 3D system sizes: 8$^3$, 12$^3$ and 16$^3$.
(b) Estimating the thermodynamic T$_{\rm c}$ using finite size scaling. M(T) for different system sizes collapses to a universal function close to 
T$_{\rm{c}}$ with universal critical exponents. We used 3D O(3) universality class exponents and $\epsilon=|{\rm T}-{\rm T}_{\rm c}|/{\rm T}_{\rm c}$ 
is the reduced temperature. 
As a result, we found T$_{\rm c}$=0.14$t$ for SFMO. 
\label{Fig:4}}
\end{figure}

\noindent {\em Phase transition and determination of T$_{\rm c}$:}
The primary advantage of the classical Hamiltonian H$_{\rm{eff}}$ is our ability to simulate much larger system 
sizes compared to those using ED+MC methods. We have performed the first 3D finite temperature simulations of 
magnetic properties using classical Monte Carlo on up to 16$^3$ unit cells on an
FCC lattice, as shown in Fig.~\ref{Fig:4}(a)).
We have determined ${\rm T_c}$ using the finite size scaling of M(T).
According to the finite size scaling hypothesis, M(T) for a system of size $L^3$ is described by 
a function of the form M(T,L)=L$^{-\beta/\nu}\mathcal{F}(\epsilon {\rm L}^{1/\nu})$ where $\mathcal{F}(x)$ is a universal 
function and $\epsilon=\lvert {\rm T}-{\rm T}_{\rm c} \rvert / {\rm T}_{\rm c}$. The critical exponents $\beta=0.36$ and $\nu=0.70$ are known for the 3D O(3) 
universality class. Using T$_{\rm c}$ as a fitting parameter, we plot ${\rm M}(\epsilon) {\rm L}^{\beta/\nu}$ against $\epsilon {\rm L}^{1/\nu}$ 
for L = 8, 12 and 16. For the true thermodynamic T$_{\rm c}$ all curves, of different system sizes, collapse onto a single curve,
as shown in Fig.~\ref{Fig:4}(b) providing an estimate of T$_{\rm c}=0.14t$ for SFMO. Comparing with the experimental T$_{\rm c}=420$ K, gives $t=0.27$ eV 
which is in good agreement with electronic structure calculations\cite{Sarma_2000}. 

\noindent {\it Low temperature spin wave contribution to M(T):}
Standard ferromagnetic spin waves produce a T$^{3/2}$ reduction of the magnetization, also known as the Bloch T$^{3/2}$ 
law\cite{Fazekas}. 
However, in Fig. \ref{Fig:3}(c), M(T) is linear at low T and this linear behavior in fact persists up to a relatively large fraction of T$_{\rm c}$. 
We explain this difference, between the Bloch Law and the calculated linear behavior, as arising from the difference between classical and quantum magnons.
The classical Hamiltonian is equivalent to taking the $S\rightarrow \infty$ limit of the quantum Hamiltonian but keeping 
T$_{\rm c} \sim JS^2$ constant. The T$^{3/2}$ law is restricted to a temperature scale $T_0 \lesssim T_c/S$, the magnon bandwidth
or equivalently to ${\rm T}_0/ {\rm T}_{\rm c} \sim 1/S$. Therefore the range of temperatures to observe the Bloch law is completely quenched in classical calculations and highly suppressed
in the experiment due to the large S=5/2 on Fe. 

In order to understand the origin of the linear temperature dependence of the magnetization, we consider the reduction in M(T) due to spin waves described by,
\begin{eqnarray}
{\rm M(T)=M_0}\Big[1-\int_{\rm{1^{st} {\rm B.Z.}}} \frac{d^{3}q}{e^{\beta JS w_q}-1} \Big]
\label{Eq:Nk}
\end{eqnarray}
where the integral is over the first Brillouin zone. For small $q$, the dispersion for magnons $w_q \sim q^2$. As $S\rightarrow \infty$ 
the exponential can be expanded at all temperatures: $ e^{\beta JS w_q}=e^{\frac{\beta T_c w_q}{S}} \approx 1+ \frac{\beta T_c w_q}{S}$
for a constant T$_{\rm c}$.
Upon using this expansion and evaluating the integral gives M(T)$\sim {\rm M}_0(1-\alpha {\rm T})$ where $\alpha = {\cal{O}} (1) $.
Thus classical spin waves indeed provide a natural explanation of the linear T dependence of the magnetization at low T.
\begin{figure}[!t]
\includegraphics[width=6cm]{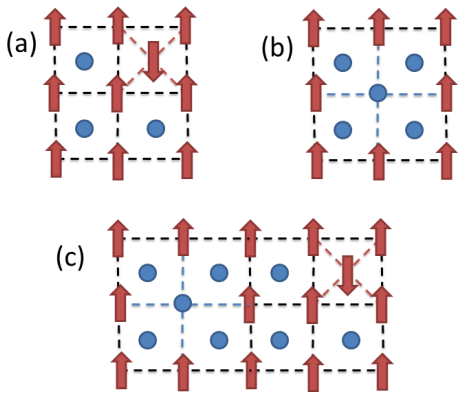}
\caption{Types of disorder: (a) excess Fe, (b) excess Mo, (c) anti-site disorder. Black, blue and red lines represent FM bonds, 
the broken FM bonds and the superexchange between Fe sites.
\label{Fig:5}}
\end{figure}

The reason for the robustness of the the linear M(T) dependence up to 
relatively high temperatures is the peculiar double square root form of H$_{\rm{eff}}$ (See Fig.~\ref{Fig:2}). Compared to the
Heisenberg Hamiltonian, H$_{\rm{eff}}$ is harmonic ($E\sim J\theta^2$) for a larger domain of $\theta$. Therefore magnon-magnon
scattering which is mainly due to the non-harmonic part of the Hamiltonian is highly suppressed and that explains 
why the spin wave regime and correspondingly the linear T behavior of M(T) survives up to relatively high T. 
Similar M(T) has been observed in experiments both on 
single crystals\cite{Tomioka_2000} and on thin films\cite{Hauser_2011}.

\section{Disorder}
In SFMO, there are three common types of disorder: excess Fe, excess Mo and anti-site disorder 
(See Fig. \ref{Fig:5}). By using H$_{\rm{eff}}$, we perform large scale calculations of the temperature-dependent magnetic properties on systems 
up to 16$^3$  to investigate the effects of disorder. Finite size effects close to T$_{\rm{c}}$  are highly
suppressed with increasing system size as shown in Fig. \ref{Fig:4}(a). 
We start the discussion with the general chemical formula Sr$_2$Fe$_{1+y}$Mo$_{1-y}$O$_6$ with 
y greater (smaller) than zero corresponding to excess Fe (Mo), followed by anti-site disorder. We conclude with a proposal to increase
T$_{\rm{c}}$ without sacrificing conduction electron polarization.

\noindent {\it{Excess Fe:}}  For $y>$0, as seen in Fig. \ref{Fig:5}(a) Fe replaces Mo sites which has two main effects: 
First, it reduces the total conduction electron density that weakens the double exchange mechanism. 
Secondly, when two Fe sites are close to each other, the strong antiferromagnetic superexchange
locks the spins. We estimate the strength of this superexchange $S(S+1)J_{\rm{AF}} \sim 34 \rm\ {meV}$
based on T$_{\rm N}$=750K for a similar compound LaFeO$_3$ with S=5/2 spins on Fe.
The excess Fe spin with the down orientation on the Mo site couples antiferromagnetically to the four neighboring up spins creating a local puddle that {\em enhances}
ferromagnetism in its neighborhood. Capitalizing on this enhanced ferromagnetism will form the basis of our proposal to enhance T$_{\rm{c}}$.

\begin{figure}[!t]
\includegraphics[width=8cm]{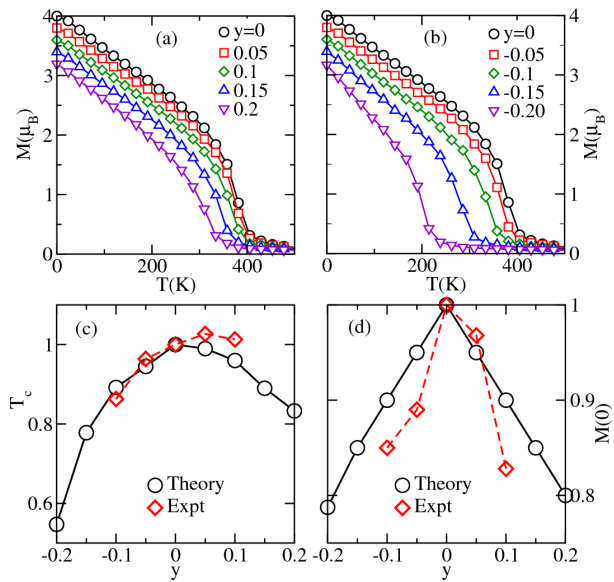}
\caption{Effects of Fe \& Mo disorder for Sr$_2$Fe$_{1+y}$Mo$_{1-y}$O$_6$ using H$_{\rm{eff}}$ and comparing it with experiments.
(a) Fe rich ($y>0$) M(T), (b) Mo rich ($y<0$) M(T), (c) T$_{\rm c}$ as a function of $y$, (d) Saturation magnetization, M(0), with $y$ 
compared with experiments\cite{Topwal_PRB2006}.
\label{Fig:6}}
\end{figure}

Fig.~\ref{Fig:6}(a) shows that the saturation magnetization M(0)
drops with increasing amount of excess Fe, largely because of its antiferromagnetic coupling to the neighboring
Fe sites (see Fig. \ref{Fig:6}(d)). For small values of $y$, T$_{\rm{c}}$ does not change significantly, then 
drops rapidly (Fig. \ref{Fig:6}(c)) beyond $y\simeq0.1$. The initial insensitivity of T$_{\rm{c}}$ on $y$ can be attributed to the two effects of excess Fe cancelling each other: 1) Reduction of conduction electrons weakens FM, 2) Formation of ferromagnetic puddles locally stabilizes FM. 
The behavior of both M(0) and T$_{\rm{c}}$ as a function of $y$ are in good agreement with experiments\cite{Topwal_PRB2006,yoshida_2011}.  

\noindent {\it{Excess Mo:}} Excess Mo ($y<0$) leads to a dilution of the ferromagnetic 
bonds (see Fig. \ref{Fig:5}) as well as an increase in conduction electron density. 
The detrimental effects of dilution and broken ferromagnetic bonds on the magnetization as a function of T 
is shown in Fig. \ref{Fig:6}(b)
and reflected directly in the rapid decrease of
saturation magnetization M(0) and T$_{\rm{c}}$ as a function of $y$ (see Fig. 
\ref{Fig:6}(d)). Once again these results are in good agreement with experiments\cite{Topwal_PRB2006}. The behavior of M(0) in 
off-stoichiometric SFMO is also in agreement with DFT calculations\cite{mishra_2010}.

\begin{figure}[!t]
\includegraphics[width=8cm]{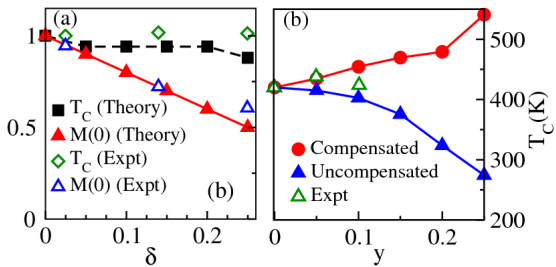}
\caption{(a) Anti-site disorder results for T$_{\rm{c}}$ and saturation magnetization, M(0) (both normalized with respect 
to their disorder-free values) compared with experiments, (b) Proposal to increase T$_{\rm{c}}$ by La and Fe doping, 
La$_x$Sr$_{2-x}$Fe$_{1+y}$Mo$_{1-y}$O$_6$. T$_{\rm{c}}$($y$) for compensated ($x=3y$) and uncompensated ($x=0$).The 
uncompensated T$_{\rm{c}}$($y$) is compared with experiments\cite{Topwal_PRB2006}.
\label{Fig:7}}
\end{figure}

\noindent {\it{Anti-site disorder:}} A realization of anti-site disorder (AS) in which Fe and  Mo sites replace each other is 
shown in Fig. \ref{Fig:5}(c). This is the most prevalent type of disorder in SFMO. It can be thought of as a combination of excess 
Fe and Mo disorder while keeping the carrier density constant. We quantify AS disorder using $\delta$ the fraction of Fe atoms 
that are on the Mo sublattice; $\delta=0.5$ is a fully disordered system. Fig. \ref{Fig:7}(a)
shows that M(0) drops linearly with a slope of ($1-2 \delta$), primarily due to the Fe spin on the wrong sublattice flipping from the parallel 
to the antiparallel direction, as shown in Fig. \ref{Fig:5}(c).
T$_{\rm{c}}$ appears to be insensitive to AS disorder, primarily because two effects balance each other. 
While the broken FM bonds in the Mo rich regions weakens FM, the puddles of Fe rich regions has the opposite effect.
Although Fe sites are coupled antiferromagnetically in these puddles, it locally creates stronger ferromagnetic domains.
We believe that these two effects balance each other and T$_{\rm{c}}$ does not change significantly with anti-site disorder, again in
very good agreement with experiments\cite{Topwal_PRB2006}.  

\noindent {\it{Proposal to increase T$_{\rm{c}}$:}} We conclude with a proposal to increase T$_{\rm{c}}$ without sacrificing conduction 
electron polarization. We propose adding excess Fe, that locally creates strong ferromagnetic puddles, and simultaneously adding extra La
to compensate the loss of carriers. Our results are shown in Fig. \ref{Fig:7}(b) and suggest that with adequate amount of La doping, 
T$_{\rm{c}}$ can be increased by about 100K.

The general formula for both La and Fe doping is La$_x$Sr$_{2-x}$Fe$_{1+y}$Mo$_{1-y}$O$_6$. Assuming that the Fe valency remains 
fixed at +3, and only Mo valency changes from +5 to +5+$\eta$ with doping, the charge balance dictates that $\eta=(2y - x)/(1 -y)$.
The corresponding carrier concentration is $n = (1 + x - 3y)/3$. This implies that setting $y=3x$ exactly compensates the lost carriers 
due to excess Fe and fixes the filling at $n=1/3$. The dependence of T$_{\rm c}$ on excess Fe for the compensated case is shown in 
Fig. \ref{Fig:7}(b). We find that T$_{\rm c}$ increases by as much as 100 K for $y=0.25$. Next we argue that our approach for enhancing 
${\rm T_c}$ is better than only La doping.  
It is known that La substitution of $x = 1$ gives rise to a 15$\%$ increase of T$_{\rm c}$\cite{Navarro_PRB2001}.
However, this is accompanied by a huge increase in the extent of anti-site disorder\cite{Navarro_PRB2001}. 
For x=1, Mo valence changes from +5 to +4 (using $\eta=-x$). The reduced electrostatic attraction
between the Mo and the surrounding oxygen octahedra leads to an expansion of the MoO$_6$
octahedra. As the volume of the MoO$_6$ octahedra approaches that of
FeO$_6$, the B-B' ordering  becomes fragile\cite{Ibarra_2007}
and the increased anti-site disorder reduces the polarization 
significantly\cite{our_PRB1}. In contrast, our
proposal suggests a 25$\%$ increase in T$_{\rm c}$ is obtained for $y = 0.25$ and $x = 0.75$, with
an average Mo valence of +4.66 which is unlikely to give rise to
large amounts of anti-site disorder.

Finally, we have checked that the proposed system with excess Fe and La compensation is indeed fully polarized at T=0 using ED+MC.
The increase in 
T$_{\rm{c}}$ by about 100K is extremely encouraging as that would increase the room temperature polarization significantly. 

\section{Conclusion}
We have found a non-trivial generalization of the double exchange mechanism that is relevant for driving ferromagnetism in 
the double perovskite half metals. The effective magnetic Hamiltonian H$_{\rm{eff}}$ with the double square root form, obtained after integrating out the itinerant electrons, 
is very different from standard Heisenberg or double exchange 
Hamiltonians and agrees remarkably well when compared with the full quantum Hamiltonian. 
H$_{\rm{eff}}$ is found to retain the harmonic $\theta^2$ form in the canting between neighboring spins up to a larger range of $\theta$.
As a result classical spin waves provide a good description of the temperature dependent M(T), with suppressed magnon-magnon scattering. 
We have performed large scale simulations of H$_{\rm{eff}}$ with different types of disorder. From our insights on the dependence of the 
saturation magnetization and T$_{\rm{c}}$ on disorder,  we propose a mechanism to substantially increase
T$_{\rm{c}}$ by balancing excess Fe doping and compensating the loss of carriers with La doping. 

\section{Acknowlegments}
We thank D. D. Sarma for fruitful discussions. Funding for this research was provided by the Center for Emergent Materials at
the Ohio State University, an NSF MRSEC (Award Number DMR-0820414). 

\appendix
\section{Including effects of $\Delta$ in H$_{\rm eff}$}
\begin{figure}[!t]
\includegraphics[width=5.5cm]{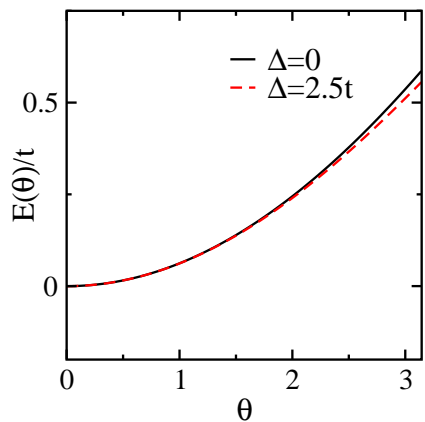}
\caption{Comparison of energy as a function of $\theta$ for the two site problem calculated using the spin Hamiltonian in eq. 4 by setting $\Delta=0$ (solid black) and $\Delta=2.5$ (dashed red). They agree well for all values of $\theta$.
\label{Fig:A1}}
\end{figure}

Here we show that the effects of $\Delta$ can be included in the spin Hamiltonian shown in eq. 10 which we derived by setting $\Delta = 0$. As justified in the main text, for small deviations from the ferromagnetic ground state the spin wave dispersion has the Heisenberg form which can be captured by appropriately fitting the spin wave spectrum of H$_{\rm eff}$ to that of the full quantum Hamiltonian with $\Delta \neq 0 $. There is, however, still the question of how well the model describes large spin canting which is the main focus of our work. In Fig.~\ref{Fig:A1} we have shown the energy as a function of $\theta$ for the two site problem calculated using the spin Hamiltonian in eq. 4 by setting $\Delta=0$ and $\Delta=2.5$. It is clearly seen that once the spin stiffness is appropriately choosen to match the low energy dispersion, the two models agree within a precission of less than 3\% for all values of $\theta$. This justifies our approach of using the simplest model with $\Delta=0$. 

\bibliography{references}
\bibliographystyle{apsrev}

\end{document}